\documentclass[lettersize,journal]{IEEEtran}
\usepackage{amsmath,amsfonts}
\usepackage{algorithmic}
\usepackage{algorithm}
\usepackage{array}
\usepackage[caption=false,font=normalsize,labelfont=sf,textfont=sf]{subfig}
\usepackage{textcomp}
\usepackage{stfloats}
\usepackage{url}
\usepackage{verbatim}
\usepackage{graphicx}
\usepackage{multirow}
\usepackage{makecell}
\usepackage{cite}
\begin{document}

\title{Secret-to-Image Reversible Transformation for Generative Steganography}

\author{Zhili Zhou,~\IEEEmembership{Member,~IEEE,} Yuecheng Su, Q. M. Jonathan Wu,~\IEEEmembership{Senior Member,~IEEE,} Zhangjie Fu, Yunqing Shi,~\IEEEmembership{Life Fellow,~IEEE,}
        % <-this % stops a space
\thanks{This work was supported in part by the National Natu-ral Science Foundation of China under Grant 61972205, U1836208, U1836110, in part by the National Key R\&D Program of China under Grant 2018YFB1003205, in part by the Priority Academic Program Development of Jiangsu Higher Education Institutions (PAPD) fund, in part by the Collaborative Innovation Center of Atmos-pheric Environment and Equipment Technology (CI-CAEET) fund, China.}% <-this % stops a space
\thanks{Z. L. Zhou, Y. C. Su, and Z. J. Fu are with School of Computer and Software \& Engineering Research Center of Digital Forensics Ministry of Education, Nanjing University of Information Science and Technology, Nanjing 210044, China. E-mail: zhou\_zhili@163.com, su\_yuecheng@nuist.edu.cn, fzj@nuist.edu.cn.}
\thanks{Q. M. J. Wu is with the Department of Electrical and Computer Engineering, University of Windsor, Windsor, Ontario, Canada N9B 3P4. E-mail: jwu@uwindsor.ca.}
\thanks{Y. Q. Shi is with the Department of ECE, New Jersey Institute of Technology, Newark, NJ 07102 USA. E-mail: shi@njit.edu.}}

\markboth{Journal of \LaTeX\ Class Files,~Vol.~14, No.~8, August~2021}%
{Shell \MakeLowercase{\textit{et al.}}: A Sample Article Using IEEEtran.cls for IEEE Journals}

\IEEEpubid{0000--0000/00\$00.00~\copyright~2021 IEEE}

\maketitle

\begin{abstract}
Recently, generative steganography that transforms secret information to a generated image has been a promising technique to resist steganalysis detection. However, due to the inefficiency and irreversibility of the secret-to-image transformation, it is hard to find a good trade-off between the information hiding capacity and extraction accuracy. To address this issue, we propose a secret-to-image reversible transformation (S2IRT) scheme for generative steganography. The proposed S2IRT scheme is based on a generative model, i.e., Glow model, which enables a bijective-mapping between latent space with multivariate Gaussian distribution and image space with a complex distribution. In the process of S2I transformation, guided by a given secret message, we construct a latent vector and then map it to a generated image by the Glow model, so that the secret message is finally transformed to the generated image. Owing to good efficiency and reversibility of S2IRT scheme, the proposed steganographic approach achieves both high hiding capacity and accurate extraction of secret message from generated image. Furthermore, a separate encoding-based S2IRT (SE-S2IRT) scheme is also proposed to improve the robustness to common image attacks. The experiments demonstrate the proposed steganographic approaches can achieve high hiding capacity (up to 4 \textit{bpp}) and accurate information extraction (almost 100\% accuracy rate) simultaneously, while maintaining desirable anti-detectability and imperceptibility.
\end{abstract}


\section{Introduction}
\IEEEPARstart{I}{mage} steganography is the technology of imperceptibly hiding secret information into a carrier image so that the covert communication can be achieved by transmitting the carrier image without arousing suspicion \cite{r1,r2}. As another way of covert communication, data encryption usually encodes the secret information in an incomprehensible and meaningless form, and thus it exposes the importance and confidentiality. That makes the secret information vulnerable to be intercepted and cracked. Therefore, compared to encryption, the advantage of steganography is that it can conceal the occurrence of covert communication to ensure the security of the transmitted secret information \cite{r1,r2}.

The conventional steganographic approaches generally select an existing image as a cover and then embed secret information into the cover with a slight modification. However, due to the modification, image distortion is inevitably left more or less in the cover image, especially under a high hiding payload. That might cause the presence of hidden information to be successfully exposed by well-designed steganalyzers \cite{r3}. To resist the detection of steganalyzers, recently, some researchers have devoted to the study of “generative steganography” \cite{r4,r5,r6}. Instead of modifying an existing cover image, the idea of generative steganography is to transform a given secret message to a generated image, which is then used as the stego-image for covert communication. Since the secret information is concealed without modification, generative steganography could achieve promising performance against the detection of steganalyzers.

Recently, generative steganography usually applied Generative Adversarial Networks (GANs) \cite{r7} to transform a given secret message to a new meaningful image that looks very realistic, i.e., “realistic-looking image”. Some generative steganographic approaches usually converted the secret message to a simple image label \cite{r8,r9,r10} or low-dimensional noise signal \cite{r6,r11,r12,r13} as the input of a GAN model to generate an image, and thus the secret message is finally transformed to the generated image. However, these approaches can only convert a secret message of small size to a low-dimensional input of GAN model, and only allow one-way mapping from the input information to the image content. Thus, the secret-to-image transformation process is generally inefficient and irreversible. Consequently, their hiding capacity is quite limited, or they cannot accurately extract the secret information from the generated image. That makes these approaches inapplicable in practice steganographic tasks.

According to the above, to realize both high-capacity hiding and accurate extraction of secret information, the key issue of generative steganography is to find an efficient and reversible transformation between secret messages and images. To this end, instead of using GANs, we explore a flow-based generative model, i.e., Glow model, which enables a bijective-mapping between high-dimensional latent vectors and images \newpage \noindent for generative steganography. By analyzing the properties of the bijective-mapping process of Glow model, we propose the secret-to-image reversible transformation (S2IRT) scheme based on Glow model to realize an efficient and reversible transformation between secret messages and generated images for information hiding and extraction.

Specifically, we encode a given secret messages as a high-dimensional latent vector in an efficient and robust manner, and then map the vector to a generated image by the Glow model for information hiding. Consequently, the secret message is finally transformed to a generated image. Also, a reverse transformation between the secret message and the generated image can be implemented for information extraction. Fig. 1 lists the framework of the S2IRT scheme for generative steganography based on Glow model. The main contributions are summarized as follows.

\begin{figure}[!t]
\centering
\includegraphics[width=3.5in]{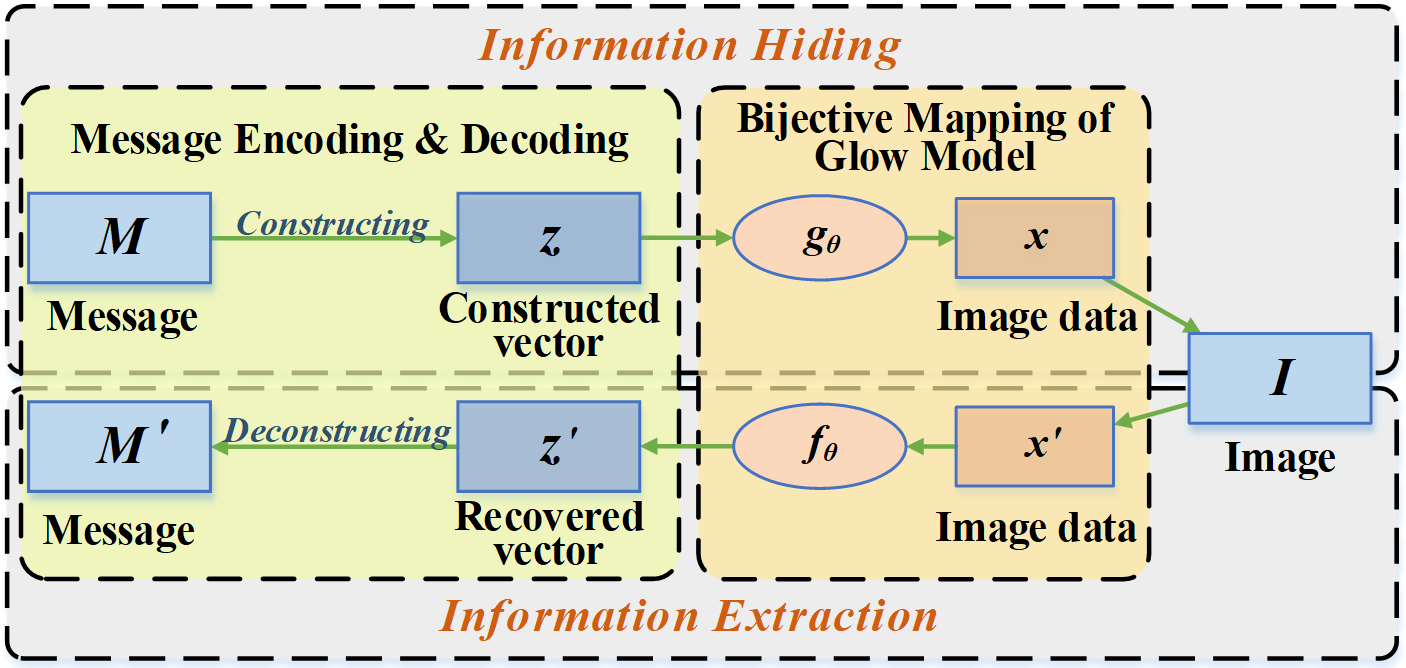}
\caption{The framework of S2IRT scheme for generative steganography based on Glow model.}
\label{fig_1}
\end{figure}

(1) The properties of the bijective-mapping process of Glow model are analyzed. By the experiments and analysis, it is found that the bijective-mapping process of Glow model has the following properties: the elements of the latent vector mapped from a real image follow one-dimensional approximately Gaussian distribution, and the mapping errors occur in the bijective-mapping process. The property analysis motivates us to propose the S2IRT scheme based on Glow model for generative steganography. 
 
(2) The S2IRT scheme is proposed for generative steganography. In the proposed S2IRT scheme, an efficient and robust encoding method is first designed to encode a given secret message as a high-dimensional latent vector. Specifically, guided by a given secret mes-sage, a large number of elements are arranged into the corresponding positions to construct a high-dimensional vector. Consequently, instead of simply encoding a given secret message as elements, the secret message is encoded as the position arrangement of these elements. The encoding method not only ensures the elements’ distributions of the constructed la-tent vector and the latent vector mapped from a real image are consistent, but also the encoded message is robust to the bijective-mapping errors. Then, the vector is mapped to a generated image by Glow model, and the generated image is used as the stego-image for steganography. The S2IRT scheme has the following advantages.

Due to the efficient and robust message encoding and the bijective mapping processes, the proposed S2IRT scheme has good efficiency and reversibility, which lead to both high hiding capacity and accurate extraction of secret message. 

Moreover, the quality of generated image can maintain at a high-level with the increase of hiding payload. That is because, although encoding secret message of different sizes lead to different constructed vectors, each constructed vector in latent space can be mapped to a high-quality image in image space by Glow model.

In addition, the existing steganalyzers are still not likely to defeat the proposed steganographic approach, since the steganography is implemented by generating a new image as the stego-image rather than by modifying an existing image.

(3) The SE-S2IRT scheme is proposed to improve the robustness of generative steganography.  To improve the robustness to common attacks such as intensity changing, contrast enhancement, and noise addition, the SE-S2IRT scheme is proposed using a designed separate encoding method for generative steganography.

The remainder of this paper is organized as follows. Section 2 introduces the related work. Section 3 briefly describes the Glow model. Section 4 and 5 elaborate the proposed S2IRT and SE-S2IRT schemes for generative steganography, respectively. Section 6 presents and discusses the experimental results. Conclusions are drawn in Section 7.

\section{Related Work}
\subsection{Conventional Steganography}
In the past two decades, a lot of conventional image steganographic approaches have been proposed. They generally select an existing image as the cover and then imperceptibly embed secret information into the cover image with a slight modification.

In the early work \cite{r14,r15}, steganography is implemented by simply modifying the least significant bit (LSB) of each pixel of cover image with the same probability to embed secret information. To reduce image distortion caused by modification, Fridrich \textit{et al.} \cite{r1} proposed the adaptive steganography based on the framework of minimizing additive distortion. Specifically, a distortion function is heuristically defined to assign an embedding cost to each pixel of the cover image, and then the secret information is embedded in the way of minimizing the sum of embedding costs, i.e., additive distortion. By designing different distortion functions, many adaptive steganographic approaches have been proposed, such as HUGO \cite{r16}, S-UNIWARD \cite{r17}, HILL \cite{r18}, and MiPOD \cite{r19}. Also, some non-additive distortion-based adaptive steganographic approaches such as DeJoin \cite{r20}, ACMP \cite{r21}, and GMRF \cite{r22} have been proposed. However, since all of these distortion functions are defined heuristically, it is still possible to design a better distortion function to implement steganography with less image distortion. 

As deep learning techniques have made remarkable achievements in the field of computer vision, some researchers used deep learning networks to automatically learn a distortion function for steganography, as done by ASDL-GAN \cite{r23}, UT-6HPF-GAN \cite{r24}, and reinforcement learning-based steganography \cite{r25}. Compared to the heuristically defined distortion functions, the distortion functions learned by deep learning networks can further reduce the image distortion for steganography. Recently, Baluja \textit{et al.} \cite{r26} trained two fully convolutional networks \cite{r27} to directly embed a secret image within a cover image and to extract the secret image from the stego-image, respectively.

As an adversary of steganography, steganalysis aims to detect the existence of hidden information in stego-images. To this end, the handcrafted steganalytic features such as first-order statistical features \cite{r28} and second-order statistics \cite{r29,r30} are extracted from the stego-images, and then are fed into the advanced ma-chine learning-based classifiers such as Support Vector Machine (SVM). Fridrich \textit{et al.} \cite{r31} proposed the Spatial Rich Model (SRM)-based steganalyzer, in which a set of high-dimensional handcrafted steganalytic features are extracted for steganalysis. In recent work \cite{r32,r33,r34}, convolutional neural networks (CNNs) were applied to automatically learn steganalytic features and classify these features for steganalysis. 

Although some conventional steganographic approaches \cite{r35,r36} could fool one or several certain steganalyzers, it is very hard for the approach to resist the detection of a set of well-designed steganalyzers. The main reason is that the modification operation of conventional steganography inevitably leads to some distortion in the stego-images, which will cause the presence of hidden information to be detected successfully.

\subsection{Generative Steganography}
Instead of modifying an existing cover image, the idea of generative steganography is to generate a new image on the basis of secret information, which is then used as the stego-image for covert communication. Since the secret information is concealed without modification, generative steganography could achieve promising performance against the detection of steganalyzers. The existing generative steganographic approach-es can be roughly categorized into two classes:  texture synthesis-based and GANs-based approaches.

Wu \textit{et al.} \cite{r4} proposed a generative steganographic approach based on reversible texture synthesis. It hides a given secret message into a generated texture image through the process of texture synthesis. Qian \textit{et al.} \cite{r37} proposed an improved version of this steganographic approach to enhance the robustness to JPEG compression. Xu \textit{et al.} \cite{r38} directly transformed the secret message to an intricate texture image for steganography. However, the texture images are generally meaningless and they somewhat resemble the encrypted data. Thus, transmitting these images on networks easily arouses attackers’ suspicion. 

Fortunately, the popular generative deep learning models, i.e., Generative Adversarial Networks (GANs) \cite{r7}, are able to produce new meaningful images that look very realistic, i.e., “realistic-looking images”. Thus, some generative steganographic approaches transformed the secret messages to the realistic-looking images based on GANs. 

Some researchers converted the secret message to a simple image label or semantic information as inter-media data, and then fed it into GANs to generate the stego-image. Liu \textit{et al.} \cite{r8} transformed a given secret message to the class label information and then fed it into the Auxiliary Classifier GAN (ACGAN) \cite{r39} to generate the stego-image. Also, a discriminator was trained to determine the class label of the generated image to extract the hidden secret message. Zhang \textit{et al.} \cite{r9} mapped the secret message to the image semantic information, and then used this information as a input condition of GANs to generate the corresponding stego-image. Qin \textit{et al.} \cite{r10} used Faster RCNN \cite{r40} to detect objects and established a mapping dictionary between binary message sequences and object labels, and then transformed the secret message to the image with corresponding label by GANs. However, since the simple label and semantic information can only contain very limited information, the hiding capacity of these generative steganographic approaches is quite low.

To enhance hiding capacity, some researchers converted the secret message to a noise signal, and then used the noise signal as the input of GANs to generate the stego-image. Hu \textit{et al.} \cite{r6,r11} directly encoded the secret message as a noise signal, and then fed it into Deep Convolutional GANs (DCGAN) \cite{r41} to generate a realistic-looking image as the stego-image. Instead of using DCGAN, Li \textit{et al.} \cite{r12} adopted Wasserstein GAN Gradient Penalty (WGAN-GP) \cite{r42} to transform the secret message to a generated image for steganography. Arifianto \textit{et al.} \cite{r13} used a word2vec model \cite{r43} to convert the secret message to a word vector,  which is then used as input vector of GANs to generate the stego-image. However, since GANs only allow one-way map-ping from the low-dimensional input information to the high-dimensional images, those GANs-based steganographic approaches cannot accurately extract secret information from these generated images, especially under high hiding payloads.

In a summary, the existing generative steganographic approaches cannot achieve a good trade-off between hiding capacity and extraction accuracy of secret message. That makes them inapplicable in practice steganographic tasks. 

To achieve both high-capacity information hiding and accurate information extraction, in this paper, we at-tempt to propose the S2IRT scheme based on Glow model to realize an efficient and reversible transformation between secret messages and generated images for information hiding and extraction. Consequently, the proposed approach can achieve high-capacity information hiding and accurate information extraction simultaneously, while maintaining desirable anti-detectability and imperceptibility for generative steganography.

\section{Glow Model}
In this paper, the proposed S2IRT and SE-S2IRT schemes are proposed relying on the flow-based generative model, i.e., Glow model for generative steganography.

To model complex distribution of high-dimensional image space, the flow-based generative models such as NICE \cite{r44}, RealNVP \cite{r45}, and Glow \cite{r46} generally learn a bijective-mapping between latent space with simple distribution and image space with complex distribution. In these models, the Glow model has superior ability of generating high quality realistic-looking natural images and enables the bijective-mapping between the latent vec-tors and the images. Thus, in this paper, we explore the Glow model to implement the S2IRT and SE-S2IRT schemes for generative steganography. The Glow model is briefly described as follows.

Let $z$ be the variable of latent space following the simple distribution $p_Z (z)$, i.e., spherical multivariate Gaussian distribution, and $x$ the variable of image space following a complex distribution $p_X (x)$. They are represented by
\begin{equation}
\label{deqn_ex1a}
z \sim p_Z(z)
\end{equation}
\begin{equation}
\label{deqn_ex1a}
x \sim p_X(z)
\end{equation}
Where, the dimension of $z$ is assumed to be the same as that of $x$, and its components $z_d$ are assumed to be independent. Thus $p_Z (z)$ can be represented by $p_Z (z)=\prod_{d}p_{Z_d}(Z_d)$, where $p_{Z_d}(z_d)$ is the one-dimensional Gaussian distribution.

\begin{figure*}[!t]
\centering
\subfloat[]{\includegraphics[width=2.3in]{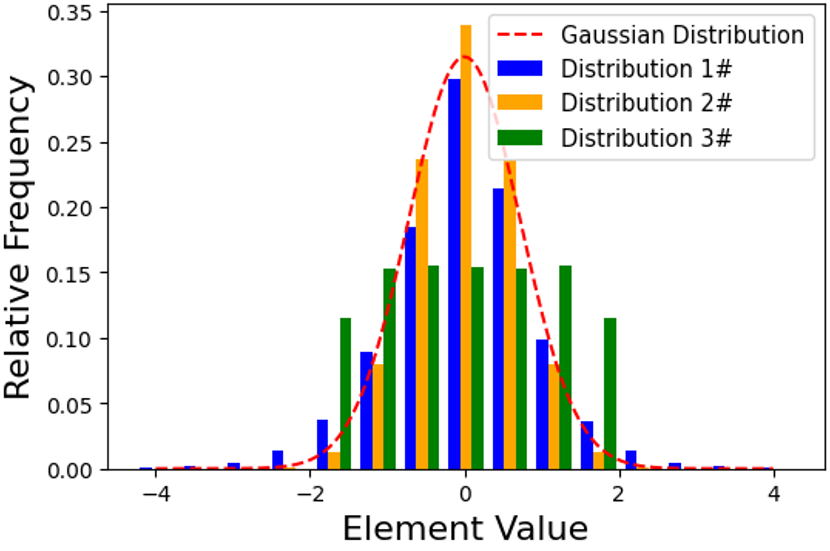}%
\label{fig_first_case}}
\hfil
\subfloat[]{\includegraphics[width=2.3in]{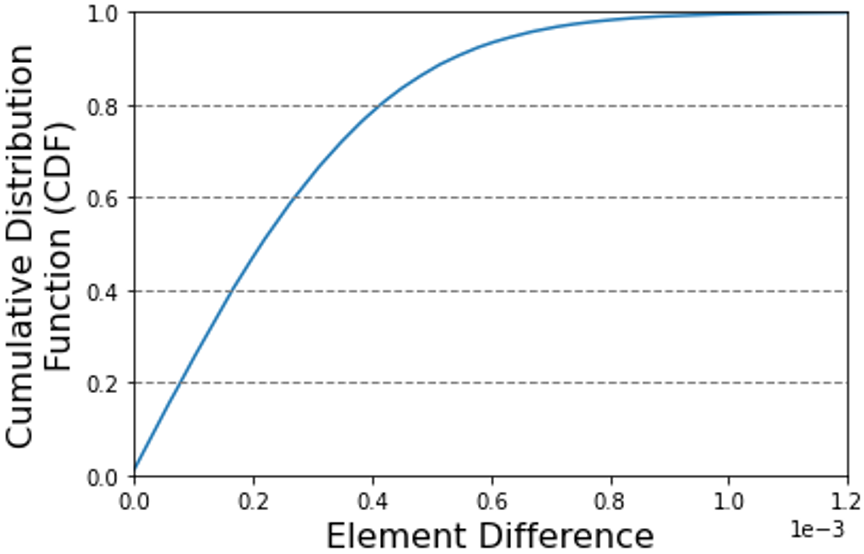}%
\label{fig_second_case}}
\hfil
\subfloat[]{\includegraphics[width=2.3in]{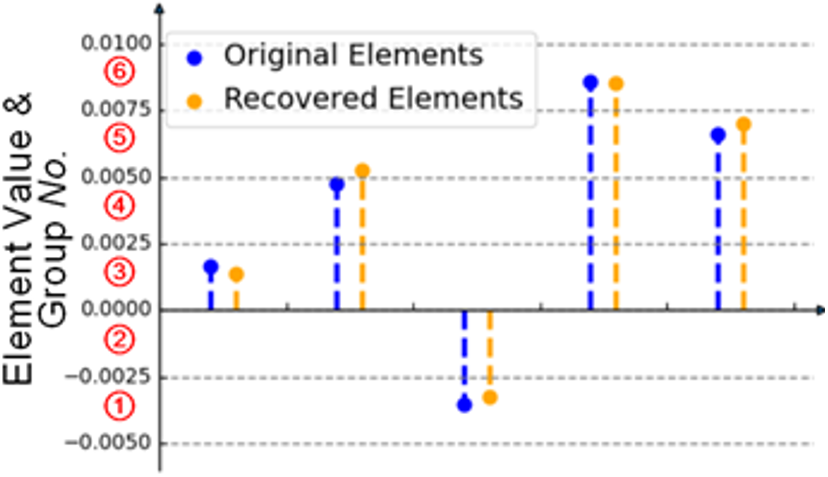}%
\label{fig_third_case}}
\caption{The illustration of the properties of the bijective mapping process in Glow model. (a) Distributions of elements of different vectors. Distribution 1\# is the distribution of the vector’s elements of a real image, Distribution 2\# the distribution of a vector constructed by element position arrangement-based encoding, and Distribution 3\# is a vector constructed by element value-based encoding. (b) The cumulative distribution function of the difference between the elements of a constructed vector and the recovered one. (c) The difference between the first five elements of the constructed and recovered vectors.}
\label{fig_sim}
\end{figure*}

To achieve the bijective-mapping, the training process of Glow model is to learn an invertible mapping function $f_\theta$. The function and its inverse function can be represented by
\begin{equation}
\label{deqn_ex1a}
z=f_\theta(x)
\end{equation}
\begin{equation}
\label{deqn_ex1a}
x=g_\theta(z)=f_\theta^{-1}(z)
\end{equation}

According to Jacobian determinant, the relationship between the distributions $p_X (x)$ and $p_Z (z)$ can be represented by
\begin{equation}
\label{deqn_ex1a}
p_X (x)=p_Z (z)\mid det\frac{\partial z}{\partial x}\mid =p_Z (f_\theta(x))\mid det\frac{\partial f_\theta(x)}{\partial x}\mid
\end{equation}
Suppose there is a training dataset $P_{data}$ containing a set of image samples $x$, and the values of their elements are in the range of $[0,1]$. According to the above equation, the learning of $f_\theta$ is realized via maximum log-likelihood estimation using the following equation:
\begin{equation}
\label{deqn_ex1a}
\begin{aligned}
&\max_{\theta \in \Theta}\mathbb{E}_{x\sim P_{data}}[\log p_X(x)] \\ 
=&\max_{\theta \in \Theta}\mathbb{E}_{x\sim P_{data}}[\log p_Z (f_\theta(x))\mid det\frac{\partial f_\theta(x)}{\partial x}\mid]
\end{aligned}
\end{equation}

Consequently, by using the learned Glow model consisting of a pair of mapping functions $f_\theta$ and $g_\theta$ as shown in Eqs. (3) and (4), we can achieve the bijective-mapping between image space and latent space.

\section{S2IRT Scheme for Generative Steganography}
Fig. 1 shows the framework of the proposed S2IRT scheme for generative steganography based on the Glow model. It consists of two stages: S2I transformation for information hiding and S2I reverse trans-formation for information extraction. In this section, we first give the motivation of the proposed S2IRT scheme, and then separately elaborate the two stages of S2IRT scheme.
\subsection{Motivation}

To realize the generative steganography, it is necessary to encode a given secret message as a latent vector at first, and then input the vector into the Glow model to generate the stego-image. To this end, we analyze the properties of bijective mapping of Glow model to pro-pose the S2IRT scheme for generative steganography. More details are given as follows.

\textbf{Property 1. The elements of the latent vector mapped from a real image follow one-dimensional approximately Gaussian distribution.}

As mentioned in Section 3, the Glow model builds a bijective-mapping between the high-dimensional latent vectors in which each dimension follows one-dimensional Gaussian distribution and the real images with a complex distribution. Moreover, the dimensions of the latent vectors are independent. Thus, for a given real image, the elements of its latent vector follow one-dimensional approximately Gaussian distribution, as shown by a toy example in Fig. 2(a). To avoid the successful identification of generated images from real images on latent space, a set of elements are randomly selected from the standard Gaussian distribution to construct a latent vector for image generation, so that the elements of the latent vector of the generated image also follow the one-dimensional approximately Gaussian distribution, as done in the literature \cite{r44,r46}.

To resist the detection of the secret message hidden in a generated stego-image on latent space, in this paper, we also randomly sample a set of elements from the standard Gaussian distribution to construct the latent vector for the stego-image generation, so that the elements of the latent vector of each generated stego-image also follow the approximately Gaussian distribution. Then, these elements are divided into a number of groups according to their values. Afterward, guided by a given secret message, we arrange the positions for these element groups sequentially to construct the latent vector, and thus the secret message is encoded as the position arrangement of these elements. 

On the contrary, if we directly encode the secret message as the element values to construct the latent vector for stego-image generation, it is very likely that the distribution of these elements is quite different from the approximately Gaussian distribution, which will cause the generated stego-image to be easily detected by analyzing the distribution of its latent vector’s elements. Fig. 2 (a) also illustrates the distributions of the elements of the latent vectors constructed by element position arrangement-based encoding and element value-based encoding methods.

\textbf{Property 2. There are mapping errors in the bijective-mapping.}
Since the bijective-mapping is implemented between continuous latent space and discrete image space, there are mapping errors in the bijective-mapping process. 

As a result, although the constructed latent vector can be recovered from the generated stego-image by the Glow model for information extraction, there is still a slight difference between the original vector and the recovered one, as shown by a toy example in Fig. 2(b). Thus, some elements of original vector near to group boundaries will shift to the adjacent groups. That will affect the accuracy of information extraction. As shown by a toy example of an original vector and the recovered one in Fig. 2(c), the second element shifts from Group 4 to Group 5, and thus it is hard to accurately obtain the position arrangement of the original vector’ elements to extract the hidden message. Thus, we choose and arrange the elements nearest to the center of each group, which are less likely to shift to the adjacent groups, for latent vector construction.
 
Therefore, motivated by above properties of bijective mapping of Glow model, we propose the S2IRT scheme for generative steganography. For information hiding, we group a set of elements randomly sampled from standard Gaussian distribution and choose the elements near to each group’s center. Then, guided by a given secret message, we arrange the positions of these elements to construct the latent vector. Finally, the constructed latent vector is input into the Glow model to generate the stego-image. Consequently, the secret message is transformed to the generated stego-image. For information extraction, the S2I reverse transformation is implemented to extract the secret message from the generated image. More details are given as follows.

\subsection{S2I Transformation for Information Hiding}
\begin{figure}[!t]
\centering
\includegraphics[width=3.5in]{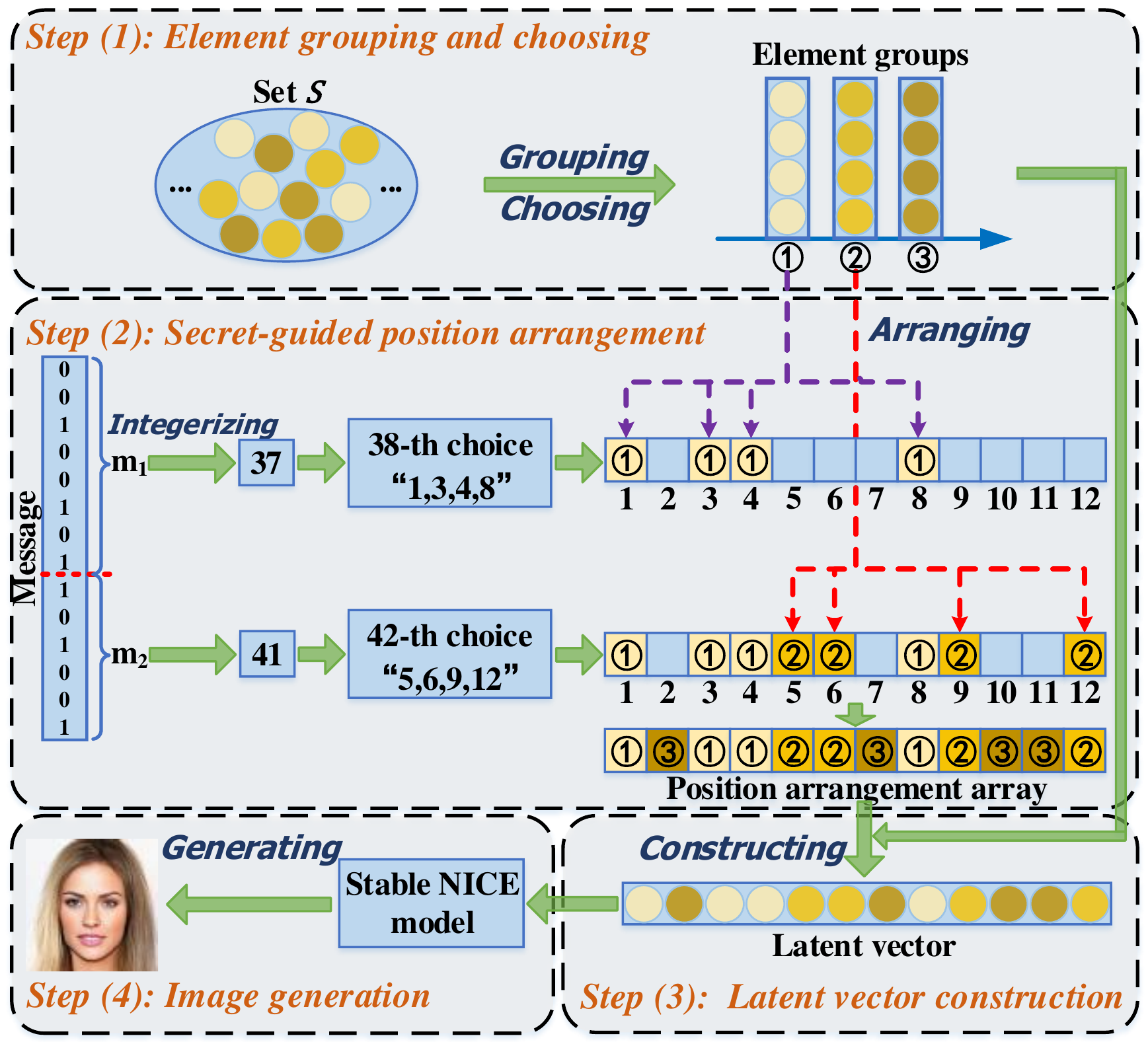}
\caption{The illustration of S2I transformation for information hiding. The process of information hiding includes four main steps: element grouping and choosing, secret-guided position arrangement, latent vector construction, and image generation.}
\label{fig_3}
\end{figure}
The S2I transformation is implemented for information hiding, as illustrated in Fig. 3. According to Section 4.1, the process of S2I transformation includes four main steps: element grouping and choosing, secret-guided position arrangement of elements, latent vector construction, and image generation, as shown in Fig. 3. These steps are detailed as follows.

\textbf{Step (1): Element grouping and choosing.} To construct a  $N_T$-dimensional latent vector for image generation, $N_T$ elements are randomly sampled from the standard Gaussian distribution at first. Then, we split the range of $[-2,2]$ into $K$ parts in the way of making the probability value of each part to be equal, as most of the sampled elements fall into this range. According to the $K$ range parts, these sampled elements are divided into $K$ groups denoted as $\{ G_i |1\le i\le K\} $.

Then, we choose $n$ elements nearest to the center of each group, which are less likely to shift to the adjacent groups, to obtain $N$ elements for latent vector construction. Thus, $N=K\times n$ and $N\le N_T$. Note that the information of $K$ and $n$ is shared between communication participants beforehand for information hiding and extraction. 

\textbf{Step (2): Secret-guided position arrangement of elements.} Guided by a given secret message $M$, the $n$ elements of each group $G_i$ are arranged into a set of corresponding $n$ positions, which are selected from the position set $pos$. Here, $pos$ records the $[N-(i-1)n]$ remaining positions for arranging these elements, where $(i-1)n$ means the number of positions occupied by the elements of $1$-\textit{th} to $(i-1)$-\textit{th} groups. The number of choices for selecting $n$ positions from $pos$ can be computed by $\omega _i=C(N-(i-1)n,n)$, where $C(x,y)$ denotes the number of choices for selecting $y$ positions from $x$ positions. Thus, $\lfloor \log_2 \omega_i \rfloor$-bit secret message can be encoded as the position arrangement of elements of $G_i$.

The secret-guided position arrangement for the elements of each group is elaborated by the pseudo code of Algorithm 1. To further illustrate Algorithm 1, we present an example of secret-guided position arrangement, which is shown in Fig. 3. Suppose there are $N=12$ elements chosen from $K=3$ groups $\{ G_i |1\le i\le 3\}$, and the number of elements chosen from each group is $n=4$. As $12$ elements needs to be arranged, the initial position set used for arranging these elements is denoted as $pos=(1,2,3,…,12)$. Guided by a given secret bitstream $M=$'00100101101001', we arrange the elements of each group into a set of positions selected from pos. More details are given as follows.

\begin{algorithm}[!t]
\caption{Secret-guided position arrangement of elements}\label{alg:alg1}
\begin{algorithmic}
\STATE
\STATE \textbf{Input:}
    \STATE \hspace{0.1cm} Secret bitstream: $M$='00100101101001'
    \STATE \hspace{0.1cm} Number of groups: $K$
    \STATE \hspace{0.1cm} Number of chosen elements in each group: $n$
    \STATE \hspace{0.1cm} Number of chosen elements in total: $N$
\STATE \textbf{Output:}
    \STATE \hspace{0.1cm} Position arrangement array: $Ind=(Ind[1],…,Ind[N])$
\STATE 
\end{algorithmic}
\begin{algorithmic}[1]
\STATE Record number of remaining positions $r \leftarrow N$
\STATE Record remaining positions $pos \leftarrow (1,2,3,…,N)$
\FOR{$i = 1$ to $K-1$}
    \STATE Compute number of choices of selecting n positions from r positions $\omega _i \leftarrow C(r,n)$
    \STATE Read the next $\lfloor\log_2\omega_i\rfloor$ bits from $M$ as a positive decimal integer $m \leftarrow $ Read($M$,$\lfloor\log_2 \omega_i \rfloor$)
    \STATE Select $n$ positions from $pos$ guided by $m$, $sel \leftarrow $Select($m$,$pos$, $n$)
    \FOR{$p$ in $sel$}
        \STATE Set $Ind[p]$ as the group \textit{No}., i.e., $i$, $Ind[p] \leftarrow i$
    \ENDFOR
    \STATE $r \leftarrow r-n$
    \STATE Remove $sel$ from $pos$, $pos \leftarrow$ PosDelete($pos$,$sel$)
\ENDFOR
\RETURN $Ind$
\end{algorithmic}
\label{alg1}
\end{algorithm}

To arrange n=4 elements of the first group $G_1$, we should select 4 positions from the 12 positions in $pos$, and thus the number of choices is $C(12,4)=495$. Since 495 choices can be used to represent $\lfloor \log_2 495\rfloor=8$ bits of secret message, we read the first 8 bits from the secret bitstream, i.e., '00100101', as the decimal integer $m=37$. Then, the $(m+1)=38$-\textit{th} position choice in lexicographical order, i.e., the positions “1,3,4,8”, are selected for arranging the elements of $G_1$. Thus, we set $Ind[1],Ind[3],Ind[4],Ind[8]$ of position arrangement array Ind as the \textit{No}. of first group, i.e., $Ind[1]=Ind[3]=Ind[4]=Ind[8]=1$. That means the positions “1,3,4,8” will be used to arrange the elements of $G_1$.

Similarly, to arrange $n=4$ elements of  $G_2$, we should select 4 positions from the remaining 8 positions in pos, and thus the number of choices is $C(8,4)=70$. The 70 choices enable the hiding of $\lfloor \log_2 70\rfloor=6$ bits of secret message. Thus, the next 6 bits are read from the secret bitstream, i.e., '101001', as the decimal integer $m=41$. Then, the $(m+1)=42$-\textit{th} position choice in lexicographical order, i.e., the positions “5,6,9,12”, are selected for arranging the four elements of $G_2$. Thus, we set $Ind[5]=Ind[6]=Ind[9]=Ind[12]=2$. That means the positions “5,6,9,12” will be used to arrange the four elements of $G_2$.

Guided by the given secret message $M$, we can finally obtain the position arrangement array $Ind=(1,3,1,1,2,2,3,1,2,3,3,2)$. Consequently, the secret message $M$ is encoded as the position arrangement of those elements. Indicated by this array $Ind$, the latent vector can be constructed by the following step.

\textbf{Step (3): Latent vector construction.} According to the obtained position arrangement array $Ind=(1,3,1,1,2,2,3,1,2,3,3,2)$, we put the four elements of $G_1$ into the “1-\textit{th}, 3-\textit{th}, 4-\textit{th}, and 8-\textit{th}” positions, the four elements of $G_2$ into the “5-\textit{th}, 6-\textit{th}, 9-\textit{th}, and 12-\textit{th}” positions, and the elements of groups $G_3$ into the remaining positions in a random order. These arranged elements are concatenated to form a $N$-dimensional vector. As there are $N_T$ sampled elements in total, the other $(N_T-N)$ elements are directly concatenated at the end of the vector to obtain the final $N_T$-dimensional latent vector $z$. 

\textbf{Step (4): Image generation.} After constructing the $N_T$-dimensional latent vector $z$, by using a private $Key$ shared between communication participants, we randomly scramble the elements of $z$, and then map the scrambled vector to a generated high-quality image by the Glow model. Finally, the generated image is used as the stego-image for covert communication.

In the proposed S2IRT scheme, the room of position arrangement of $N$ chosen elements is very large for constructing the latent vector and the group \textit{Nos}. of these elements are stable, which lead to high efficiency and robustness for encoding the secret message as the constructed vector. Moreover, the bijective-mapping between the constructed vector and the generated image can be realized by the Glow model. Thus, the proposed S2IRT scheme has good efficiency and reversibility, which lead to both high hiding capacity and accurate extraction of secret message for steganography. In addition, the constructed vector can be mapped to a high-quality image by the Glow model and the information hiding is implemented by generating a new image rather than modifying an existing image. Therefore, the proposed steganographic approach also has desirable anti-detectability and imperceptibility. These are also proven by experiments in Section 6.

\subsection{S2I Reverse Transformation for Information Extraction}
As information extraction is the reverse process of information hiding, the S2I reverse transformation is implemented to extract the secret message from the generated image. It includes two main steps:  latent vector recovery and secret message extraction.

\textbf{Step (1): Latent vector recovery.} At the receiving end, the received image is reversely mapped to a latent vector by the Glow model. Then, by the shared $Key$, the position order of the elements of the latent vector is restored to obtain the recovered latent vector $z'$.

\textbf{Step (2): Secret message extraction.} By the shared information of $K$ and $n$, the first $N=Kn$ elements of the recovered latent vector $z'$ are first re-grouped into $K$ groups in the same manner as described in the stage of information hiding.

For $n$ elements of each group $G_i$, we obtain the set of their positions in $z'$ and compute the choice \textit{No}. of this position set in all $C(N-(i-1)n,n)$ possible choices. According to step (2) of information hiding stage, the choice \textit{No}. is equal to $(m_i+1)$, where $m_i$ is the decimal integer of corresponding secret bits. Thus, for each group $G_i$, we can obtain $m_i$ from the position choice \textit{No}. of its $n$ elements and then transform $m_i$ to the corresponding secret bits. After obtaining all the bits, we concatenate these bits to extract the final secret message $M'$.

\subsection{Hiding Capacity Analysis}
As mentioned in the above subsection, for \textit{i}-th group, $\lfloor \log_2 \omega_i \rfloor$-bit secret message can be encoded as the position arrangement of its elements, where $\omega_i=C(N-(i-1)n,n)$. Thus, we can compute the total number of bits that can be encoded as the position arrangement for all groups by 
\begin{equation}
\label{deqn_ex1a}
BN_{S2I}=\sum^{K-1}_{i=1}\lfloor \log_2 \omega_i\rfloor
\end{equation}
It satisfies the following equation.
\begin{equation}
\label{deqn_ex1a}
(\sum^{K-1}_{i=1}\log_2 \omega_i)-K+1 <\sum^{K-1}_{i=1}\lfloor \log_2 \omega_i\rfloor\le \sum^{K-1}_{i=1}\log_2 \omega_i
\end{equation}
Where,
\begin{equation}
\label{deqn_ex1a}
\sum^{K-1}_{i=1}\log_2 \omega_i = \log_2 \prod^{K-1}_{i=1} C(N-(i-1)n,n)
\end{equation}
According to the Stirling's approximation, we can obtain the following result.
\begin{equation}
\label{deqn_ex1a}
\begin{aligned}
&\sum^{K-1}_{i=1}\log_2 \omega_i  \approx \log_2((\frac{1}{\sqrt{2\pi}})^{K-1}\frac{(Kn)^{Kn+1/2}}{n^{K(n+1/2)}}) \\
&=-(K-1)\log_2 \sqrt{2\pi}+(Kn+1/2) \log_2 K-\frac{K-1}{2}\log_2 n
\end{aligned}
\end{equation}

Suppose all the $N_T$  sampled elements are chosen and divided into $N_T$  groups, i.e., $N=N_T$ and $K=N_T$, and each group contains only $n=1$ element for information hiding. The sizes of the training images used in the experiments are $256\times256\times3=196,608$ and the dimension of constructed latent vector, i.e., $N_T$, is the equal to the sizes of those images, and thus $N_T=196,608$. According to Eq. (8) and (10), the maximum $BN_{S2I}$ is approximately estimated to be in the range of $(3,000,094, 3,196,701]$, and thus the maximum number of bits hidden in per image pixel per channel (bpp) can theoretically reach up to a very high value, i.e., $3,196,701/196,608\approx 16.2$ \textit{bpp}. 

According to Eq. (10), larger $K$ and $n$ lead to higher capacity of information hiding but lower accuracy of information extraction, which is also analyzed in Section 6.2.

\section{SE-S2IRT Scheme For Generative Steganography}
\begin{figure}[!t]
\centering
\includegraphics[width=3.5in]{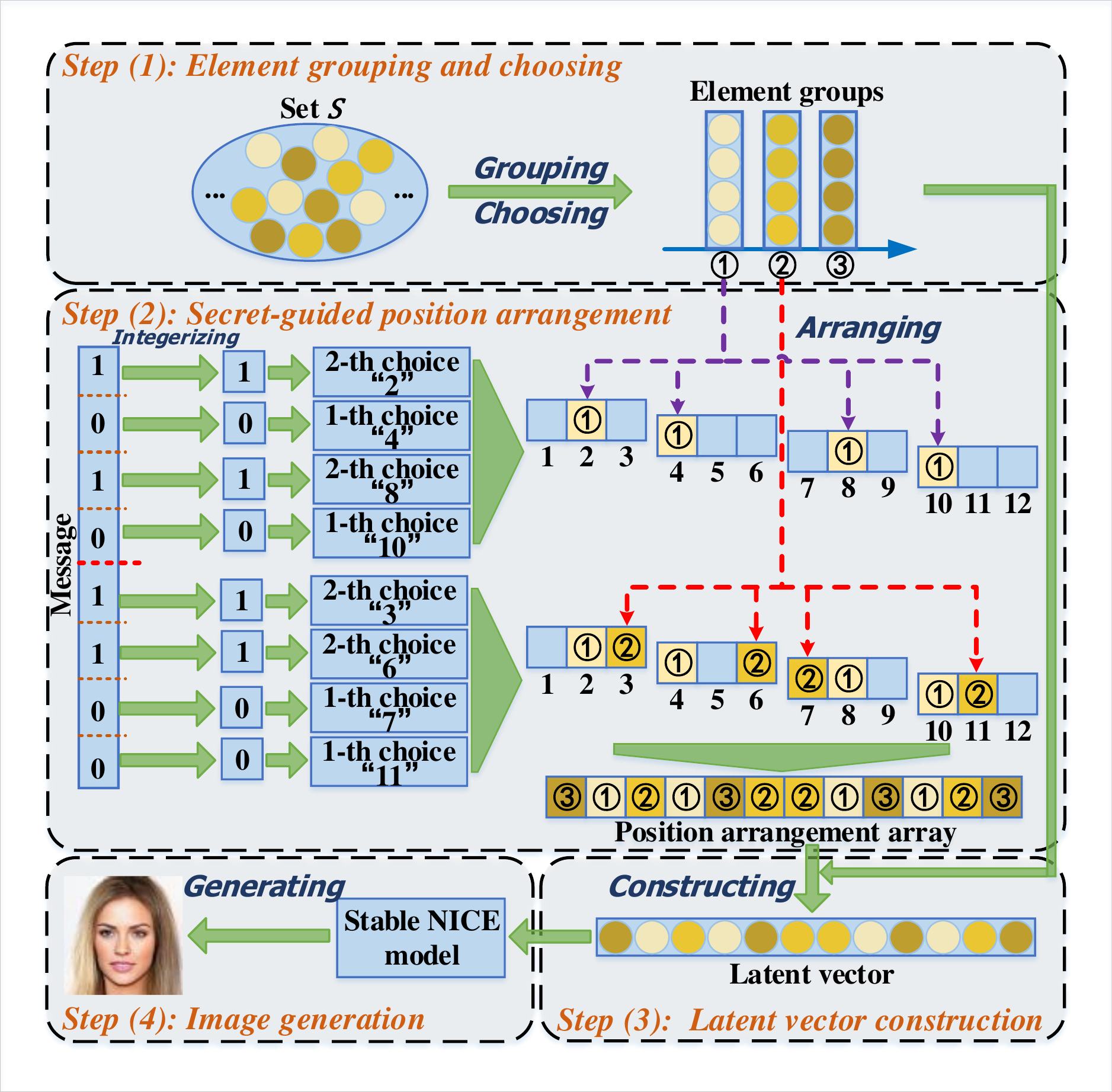}
\caption{The illustration of SE-S2I transformation for information hiding. The process of information hiding is mainly divided into four steps: element grouping and choosing, secret-guided position arrangement, latent vector construction, and image generation.}
\label{fig_4}
\end{figure}
In practice, images usually suffer from a variety of attacks such as intensity changing, contrast enhancement and noise addition during transmitting on networks. In the S2IRT scheme, although the group \textit{Nos}. of the chosen elements used for latent vector construction are relatively stable, those attacks will cause some elements of the constructed latent vector to shift from one group to its adjacent ones. Once the group \textit{No}. of one element is changed, the position choice \textit{Nos}. for arranging the elements of most groups would be changed and thus most secret bits encoded as the position arrangement for these groups will be affected. Consequently, it is hard for the S2IRT scheme to accurately extract secret information from the recovered latent vector under those attacks. To improve the robustness to those attacks, we propose a separate encoding-based S2IRT (SE-S2IRT) scheme for steganography.

\subsection{SE-S2I Transformation for Information Hiding}
The proposed SE-S2IRT scheme is based on the idea of separate encoding of secret message. Fig. 4 illustrates the process of SE-S2I transformation for information hiding, which also includes following four steps.

\textbf{Step (1): Element grouping and choosing.} This step is the same as that of S2IRT scheme. The $N_T$ sampled elements are randomly sampled from the standard Gaussian distribution and divided into K groups denoted as $\{ G_i |1\le i\le K\}$. Then, $n$ elements nearest to the center of each group are chosen to obtain the $N$ elements, and thus $N=K\times n$ and $N\le N_T$.

\textbf{Step (2): Secret-guided position arrangement of elements.} To improve the robustness to image attacks, guided by the secret message $M$, for each group $G_i$, we select one position from every $[N-(i-1)n]/n=(K-i+1)$ adjacent positions of the position set pos to separately arrange the $n$ elements of $G_i$. Where, pos records the $[N-(i-1)n]$ remaining positions, and $(i-1)n$ means the number of positions occupied by the elements of 1-\textit{th} to $(i-1)$-\textit{th} groups. The number of choices for selecting one position from $(K-i+1)$ positions can be computed by $\omega_i=C(K-i+1,1)$, and thus $\lfloor \log_2 \omega_i \rfloor$ bits can be encoded as the position arrangement for one element of $G_i$. As there are $n$ elements in $G_i$, totally $n\lfloor \log_2 \omega_i \rfloor$ can be encoded as the position arrangement for the n elements of $G_i$.

To illustrate the position arrangement, Fig. 4 shows an example of secret-guided position arrangement using the SE-S2IRT scheme. Suppose there are totally $N=12$ chosen elements in $K=3$ groups $\{G_i |1\le i\le 3\}$ and the number of elements in each group is $n=4$. As 12 elements need to be arranged, the initial position set used for arranging these elements is $pos=(1,2,3,…,12)$. Then, guided by a given secret bitstream $M$='10101100', we separately arrange the elements of each group into the corresponding selected positions. More details are given as follows.

For arranging one element of $G_1$, the number of choices of selecting one position from 3 adjacent positions is $C(3,1)=3$, and thus $\lfloor \log_2 3=1$ can be encoded. Therefore, we read the first one bit from the secret bitstream, i.e., '1', as the decimal integer $m=1$. Then, we select the $(m+1)=2$-\textit{th} position choice in lexicographical order, i.e., the position '2', to arrange one element of $G_1$. Next, we set the $Ind[2]$  of position arrangement array Ind as the \textit{No}. of first group, i.e., $Ind[2]=1$. Similarly, we read the next bit, i.e., '0', as the decimal integer $m_2=0$, and then use the $(m+1)=1$-\textit{th} position choice in the next three adjacent positions, i.e., the position '4', to arrange one element of $G_1$. Thus,  $Ind[4]=1$. After separately arranging all the 4 elements of $G_1$ in the same manner, totally 4 bits, i.e., '1010', are encoded, and $Ind[2]$, $Ind[4]$, $Ind[8]$ and $Ind[10]$ are set as 1. The element arrangement process is also shown in Fig. 4.

The position arrangement for other groups is the same as that for $G_1$. Guided by the given secret message $M$, finally, we can select the corresponding positions for the elements of all groups and obtain the position arrangement array $Ind=(3,1,2,1,3,2,2,1,3,1,2,3)$. The secret message $M$ is encoded as as the position arrangement of those elements, and Ind will be used for latent vector construction.

\textbf{Step (3): Latent vector construction} and \textbf{Step (4): Image generation} are also the same as those of S2IRT scheme. Thus, they are skipped. 
In the SE-S2IRT scheme, once the group \textit{No}. of one element arranged into one of $K$ adjacent positions is changed, only the position choice \textit{Nos}. for arranging the elements into the $K$ adjacent positions will be changed and thus only $\sum^{K-1}_{i=1} \lfloor\log_2 (K-i+1)\rfloor$ bits will be affected as most. Therefore, the SE-S2IRT scheme provides better robustness to common image attacks than the S2IRT scheme, which is also proven by experiments in Section 6.3.

\subsection{SE-S2I Reverse Transformation for Information Extraction}
Similarly, the reverse SE-S2I transformation is implemented for information extraction.

\textbf{Step (1): Latent vector recovery.} This step is the same as that of S2IRT scheme. Thus, it is also skipped.

\textbf{Step (2): Secret message extraction.} By the shared information of $K$ and $n$, we re-group the first $N=Kn$ elements of the recovered latent vector $z'$ in the same manner as described in the stage of information hiding of S2IRT scheme.
For the $n$ elements of each group $G_i$, we obtain their positions in $z'$. Then, we compute the \textit{No}. of position choice of selecting one position from every $(K-i+1)$ adjacent positions. According to the step (2) of information hiding stage in SE-S2IRT scheme, the No. of position choice is equal to $(m_i+1)$, where $m_i$ is the decimal integer of corresponding secret bits. Then, the decimal $m_i$ is transformed to the $\lfloor \log_2 C(L-i+1,1)\rfloor$ bits. According to the position choices for the $n$ elements of each group $G_i$, we can obtain $n\lfloor \log_2 C(L-i+1,1)\rfloor$ bits of secret message. After obtaining all bits from the element positions of all groups, we concatenate these bits to extract the final secret message $M'$.

\subsection{Hiding Capacity Analysis}
As mentioned in the above subsection, for i-th group, $n\lfloor \log_2 \omega_i\rfloor$-bit secret message can be encoded as the position arrangement of its elements, where $\omega_i=C(K-i+1,1)$. Thus, we can compute the total number of bits encoded as the position arrangement for all groups by
\begin{equation}
\label{deqn_ex1a}
BN_{SE-S2I}=\sum^{K-1}_{i=1}n\lfloor \log_2 \omega_i\rfloor
\end{equation}
It satisfies the following equation.
\begin{equation}
\label{deqn_ex1a}
(\sum^{K-1}_{i=1}n\log_2 \omega_i)-n <\sum^{K-1}_{i=1}n\lfloor \log_2 \omega_i\rfloor\le \sum^{K-1}_{i=1}n\log_2 \omega_i
\end{equation}
Where,
\begin{equation}
\label{deqn_ex1a}
\sum^{K-1}_{i=1}n\log_2 \omega_i = n\log_2 \prod^{K-1}_{i=1} (K-i+1))=n\log_2 (K!)
\end{equation}
According to the Stirling's approximation, we can obtain the following result.
\begin{equation}
\label{deqn_ex1a}
\begin{aligned}
\sum^{K-1}_{i=1}n\log_2 \omega_i  &\approx n\log_2(\sqrt{2\pi K}(\frac{K}{n})^K) \\&=Kn\log_2 \frac{\sqrt[2K]{2\pi K}K}{n}
\end{aligned}
\end{equation}
Suppose all the $N_T$  sampled elements are chosen and divided into $N_T$  groups, i.e., $N=N_T$ and $K=N_T$, and each group contains only $n=1$ element. The sizes of the training images used in the experiments are $256\times 256\times 3=196,608$ and thus $N_T=196,608$. According to the Eq. (12) and (14), the maximum $BN_{SE-S2I}$ is approximately estimated to be in the range of $(2,977,102, 3,173,709]$, and thus the maximum number of bits hidden in per image pixel can theoretically reach up to a very high value, i.e., about $3,173,709/196,608\approx 16.1$ \textit{bpp}.

According to Eq. (14), larger $K$ and $n$ lead to higher capacity of information hiding but lower accuracy of information extraction in the SE-S2IRT scheme. The impacts of parameters $K$ and $n$ are also analyzed in Section 6.2.

\section{EXPERIMENTS}
In this section, we first introduce the experimental set-tings. Then, the impacts of parameters of S2IRT and SE-S2IRT schemes—the number of element groups $K$ and the number of elements $n$ chosen from each group—are analyzed and discussed. Finally, the performances of proposed stenographic approaches are tested in the aspects of information extraction accuracy, anti-detectability, imperceptibility and robustness and are compared with those of state-of-the-arts.
\subsection{Experiment Settings}
\subsubsection{Training datasets}
In \cite{r46}, the original Glow model trained on CelebA-HQ dataset \cite{r47} can effectively and efficiently produce high-quality “realistic-looking image” images. Thus, in our experiments, we also adopt this dataset to train the Glow model. The CelebA-HQ dataset contains 30,000 high-resolution face images selected from the CelebA dataset. The sizes of the images in the dataset are up to $1024\times 1024$. For the sake of training efficiency, we rescale the sizes of images to $256\times 256$ and train the Glow model on the rescaled images.
\subsubsection{Model training details}
The Glow model is trained by maximizing the objective function defined in Eq. (6) with the learning rate 0.001. The depth of each flow is set as 32 and the level of the multi-scale architecture is set as 6. After 2,000,000 iterations on the dataset, the trained Glow model is obtained.
\subsubsection{Evaluation criteria}
In the experiments, we adopt the following evaluation criteria to evaluate the hiding capacity, extraction accuracy, anti-detectability, and imperceptibility of different steganographic approaches.
\paragraph{Hiding capacity} The information hiding capacities of most image steganographic approaches are evaluated by bits per pixel (\textit{bpp}), which means the number of secret bits hidden in per pixel per channel of an image. Also, we evaluate the hiding capacity by \textit{bpp}, which is defined by
\begin{equation}
\label{deqn_ex1a}
bpp=\frac{N_T}{W\times H \times C}
\end{equation}
Where $N_T$ represents the total number of hidden secret bits, $W$, $H$, and $C$ the width, the height, and the number of channels of the image, respectively. 

\paragraph{Extraction accuracy} Due to the hiding manner of the proposed steganographic approaches, the length of original secret bitstream may be different from that of the extracted secret bitstream. Thus, the accuracy of information extraction is evaluated based on the Edit Distance (ED) between the original secret bitstream $M$ and the extracted secret bitstream $M'$. The accuracy rate is computed by
\begin{equation}
\label{deqn_ex1a}
IE_A = 1-]\frac{ED(M,M')}{\max[Len(M),Len(M')]}
\end{equation}
Where, $Len(M)$ and $Len(M')$ mean the lengths of bitstream $M$ and $M'$, respectively.

\paragraph{Anti-detectability} We evaluate the anti-detectability performance against a steganalyzer using the following detection error rate.
\begin{equation}
\label{deqn_ex1a}
P_E = \min_{P_{FA}}\frac{1}{2}(P_{FA}+P_{MD})
\end{equation}
where $P_{FA}$ is the false-alarm (FA) probability of steganalyzer and $P_{MD}$ is the missed-detection (MD) probability of steganalyzer. Larger $P_{E}$ means higher anti-detectability performances of steganographic approaches against the steganalyzer.

\paragraph{Imperceptibility} To make the hidden message imperceptible, there should be no visible difference between the qualities of the images with and without hidden information. Without the need for an existing cover image, the proposed steganographic approaches directly generate a new image as the stego-image for information hiding. Therefore, the traditional reference image quality evaluation strategies, such as RSNR and SSIM, cannot be used to test the imperceptibility of proposed steganographic approach. Thus, we adopt the no-reference image quality evaluation strategies, i.e., Blind/Referenceless Image Spatial QUality Evaluator (BRISQUE) \cite{r48} and dimensionality reduction strategy, i.e., t-distributed Stochastic Neighbor Embedding (t-SNE) \cite{r49}, to observe the imperceptibility of proposed steganographic approaches. BRISQUE is a natural scene statistic-based distortion-generic blind/referenceless image quality assessment model that implements in the spatial domain. It is trained on the TID2008 dataset (TID2008), which contains a large number of natural images distorted with various attacks, and the quality of each image is scored by people. By inputting an image into the trained model, the score of the image can be scored automatically. T-SNE is a statistical method for visualizing high-dimensional data by transforming each data to a data point in two-dimensional space.

\subsubsection{Evaluation environment}
All the experiments are conducted on an NVIDIA RTX 3090 GPU platform using PyTorch with Python interface.

\subsection{Parameter Impacts}
\begin{figure}[!t]
\centering
\includegraphics[width=3.5in]{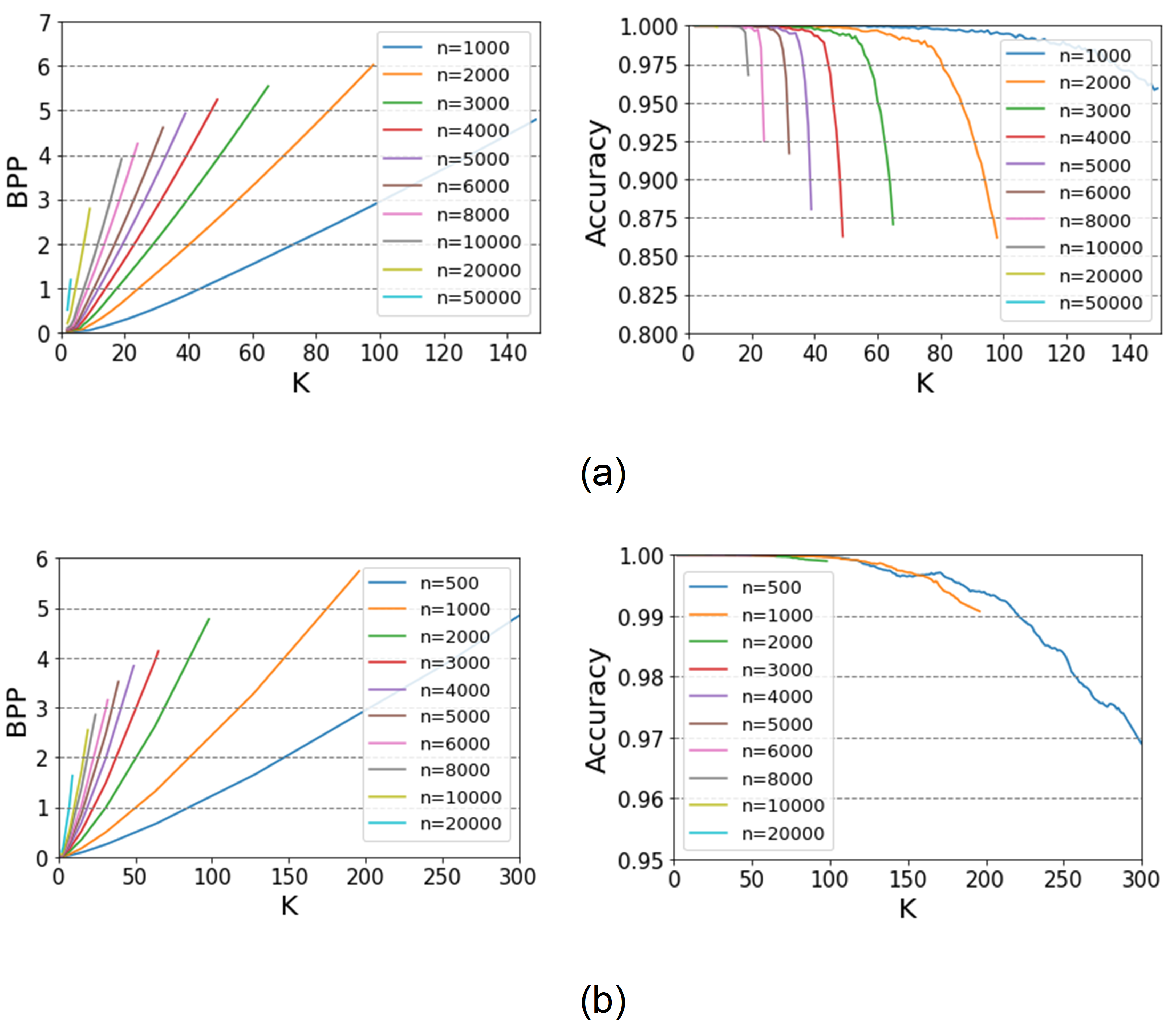}
\caption{The impacts of parameters K and n in the aspects of hiding capacity and extraction accuracy for (a) S2IRT scheme and (b) SE-S2IRT scheme.}
\label{fig_5}
\end{figure}
In the proposed S2IRT and SE-S2IRT schemes, for the latent vector construction, there are two key parameters: $K$, which is the number of element groups, and $n$, which is the number of elements chosen from each group. We test the parameter impacts on the performances of proposed schemes in the aspects of hiding capacity and extraction accuracy. In this experiment, we adopt the trained Glow model, which is introduced in previous subsection. 

Fig. 5. The impacts of parameters $K$ and $n$ in the aspects of hiding capacity and extraction accuracy for (a) S2IRT scheme and (b) SE-S2IRT scheme.
The parameter impacts on the proposed S2IRT and SE-S2IRT schemes are shown in Fig. 5(a) and (b), respectively. From the two figures, it is clear that larger $K$ and $n$ lead to higher hiding capacity. That is because larger $K$ and $n$ offer larger room of position arrangement of elements for latent vector construction, which enables more secret bits to be hidden in the image generated from the constructed latent vector. However, the accuracy of information extraction decreases with the increase of $K$ and $n$ for the following reason. Increasing $K$ and $n$ causes more elements to be near to the group boundaries, and thus the group \textit{Nos}. of these elements are vulnerable. Therefore, the group \textit{Nos}. of elements in the recovered latent vector would be quite different from those in the original latent vector, which affects the accuracy of information extraction significantly. Moreover, the hiding capacity of S2IRT is generally higher than that of SE-S2IRT at the same values of $K$ and $n$. The main reason is that the number of ways of selecting $n$ positions from $N$ positions is larger than that of selecting n positions from every $K$ positions of $N$ ones, and thus more bits can be hidden.

According to Fig. 5(a) and (b), under the condition of perfect accuracy rate of information extraction, i.e., $IE_A=1.0$, the highest hiding capacity of S2IRT and that of SE-S2IRT can reach up to 4.3 bpp and 4.1 bpp, respectively. From Fig. 5 (a) and (b), to achieve a required level of hiding capacity $IH_C$, several combinations of $(K,n)$ can be chosen. For example, when the hiding capacity is required to be $IH_C=4$ bpp for S2IRT, the set of combinations of $(K,n)$ are $(23,8000)$, $(29,6000)$, $(33,5000)$, $(40,4000)$, $(50,3000)$, $(70,2000)$ and $(128,1000)$. Also, it is found that smaller $K$ generally leads to higher extraction accuracy at a certain level of hiding capacity. Thus, the parameter combination with smallest $K$, i.e., $(23,8000)$, is chosen to achieve the required hiding capacity $IH_C=4.0$ \textit{bpp}. In the above manner, with a required level of hiding capacity, the parameter combinations $(K,n)$ can be determined for S2IRT and SE-S2IRT, and they are used in the following experiments.

\subsection{Performance Evaluation and Comparison}
In this subsection, we evaluate the performances of the proposed approaches in the aspects of information extraction accuracy, anti-detectability, imperceptibility and robustness under different hiding payloads, and compare with those of the state-of-the-arts. These approaches are listed as follows.

S-UNIWARD \cite{r17}: It is a famous conventional steganographic approach, which is based on a heuristically defined distortion function.

UT-6HPF-GAN \cite{r24}: This conventional steganographic approach is based on the distortion function learned by GAN.

Deep-Stego \cite{r26}: This conventional approach directly trains two fully convolutional networks to hide a secret image within a cover image and to extract the hidden image from the cover, respectively. 

SWE \cite{r6}: This is a generative steganographic approach, called as steganography without embedding (SWE), in which the secret information is directly encoded as the input noise of DCGAN to generate the stego-images. 

S2IRT: This is the steganographic approach using S2IRT scheme proposed in Section 4. 

SE-S2IRT: This is the steganographic approach using SE-S2IRT scheme proposed in Section 5. In both S2IRT and SE-S2IRT, the Glow model trained on the CelebA-HQ dataset is used to generate the stego-images.

\begin{table}[]
\centering
\caption{The Information Extraction Accuracy of Those Approaches with Different Hiding Payloads}
\setlength{\tabcolsep}{3mm}
\begin{tabular}{cccccc}
    \hline
    \multirow{2}{*}{Approaches} & \multicolumn{5}{c}{Hiding payloads (\textit{bpp})} \\ 
    \cline{2-6}
    & 0.1 & 0.5 & 1.0 & 2.0 & 4.0 \\
    \hline
    Deep-Stego & - & - & 0.9643 & 0.8315 & 0.7524 \\
    SWE & 0.7314 & 0.7167 & 0.7134 & 0.7120 & 0.7122 \\
    S2IRT & 1.0000 & 1.0000 & 1.0000 & 1.0000 & 0.9943 \\
    SE-S2IRT & 1.0000 & 1.0000 & 1.0000 & 1.0000 & 1.0000 \\
    \hline
\end{tabular}
\label{table_1}
\end{table}

\subsubsection{Information extraction accuracy}
Table I lists the performances of information extraction accuracy ($IE_A$) of different steganographic approaches, i.e., Deep-Stego, SWE, S2IRT, and SE-S2IRT, with the increase of hiding payloads. From this table, it is clear that the proposed steganographic approaches, i.e., S2IRT and SE-S2IRT, achieve much higher information extraction accuracy than the conventional steganographic approach, i.e., Deep-Stego, and the generative steganographic approach, i.e., SWE, under different hiding payloads. The extraction accuracy rates of S2IRT and SE-S2IRT keep at a very high-level ($IE_A\approx 1.0$) when the hiding payload ranges from 0.1 \textit{bpp} to 4 \textit{bpp}. That is because S2IRT and SE-S2IRT achieve secret-to-image reversible transformation, mainly due to the efficient and robust message encoding and the bijective mapping between latent space and image space by the Glow model. In a summary, the proposed generative steganographic approaches can achieve high hiding capacity (up to 4 \textit{bpp}) and accurate extraction of secret message (almost 100\% accuracy rate), simultaneously. 

In contrast, with the increase of hiding payload, the information extraction accuracies of Deep-Stego and SWE decrease for the following reasons. For Deep-Stego, the secret message is compressed and then is hidden into a cover image by training a fully convolutional network in a lossless way, and thus it is hard to accurately extract the secret message from the stego-image. For SWE, the secret message is encoded as a low-dimensional noise signal of DCGAN to generate the stego-image. However, DCGAN only allows one-way mapping from the input information to the image, which makes the hidden secret message difficult to be extracted from the stego-image. Thus, Deep-Stego and SWE cannot achieve accurate information extraction under high hiding payloads.

\subsubsection{Anti-detectability}
To evaluate and compare the anti-detectability performances of different steganographic approaches, we use the famous steganalyzers including SRM \cite{r32} and XuNet \cite{r33} to detect the presence of hidden information in the stego-images. Where, SRM is based on a set of high-dimensional hand-crafted steganalytic features, while XuNet is based on a modified CNN structure for steganalysis.

For the conventional steganographic approaches, the training dataset of the steganalyzers includes 5000 natural images (cover images) and 5000 stego-images obtained by embedding the secret messages into these natural images. It is notable that the generative steganographic approaches generate the stego-images by the secret-to-image transformation without the need of the natural images (cover images). For the generative steganographic approaches, as both the natural images and the generated images without information hiding can be regarded as cover images \cite{r51}, the training dataset of the steganalyzers consists of 5000 images without information hiding (including 2500 natural images and 2500 generated images without information hiding) and 5000 generated images with information hiding. By using the two trained steganalyzers, the detection error rate $P_E$ is computed in the manner described in Section 6.1 to evaluate the anti-detectability performances of different steganographic approaches.

\begin{table*}[]
\centering
\caption{The Values of $P_E$ of Steganographic Approaches with Different Hiding Payloads}
\setlength{\tabcolsep}{8mm}
\begin{tabular}{ccccccc}
    \hline
    & \multirow{2}{*}{Approaches} & \multicolumn{5}{c}{Hiding payloads (\textit{bpp})} \\ 
    \cline{3-7}
    & & 0.1 & 0.5 & 1.0 & 2.0 & 4.0 \\
    \hline
    \multirow{6}{*}{SRM} & S-UNIWARD & 0.4229 & 0.1824 & 0.0493 & - & - \\
    & UT-6HPF-GAN & 0.4414 & 0.2489 & 0.0615 & - & - \\
    & Deep-Stego & - & - & 0.0000 & 0.0000 & 0.0000 \\
    & SWE & 0.2687 & - & - & - & - \\
    & S2IRT & 0.2707 & 0.2711 & 0.2698 & 0.2702 & 0.2701 \\
    & SE-S2IRT & 0.2696 & 0.2700 & 0.2710 & 0.2703 & 0.2694 \\
    \hline
    \multirow{6}{*}{XuNet} & S-UNIWARD & 0.4461 & 0.1925 & 0.0712 & - & - \\
    & UT-6HPF-GAN & 0.4690 & 0.2971 & 0.0787 & - & - \\
    & Deep-Stego & - & - & 0.0000 & 0.0000 & 0.0000 \\
    & SWE & 0.2691 & - & - & - & - \\
    & S2IRT & 0.2701 & 0.2697 & 0.2686 & 0.2708 & 0.2702 \\
    & SE-S2IRT & 0.2701 & 0.2697 & 0.2686 & 0.2708 & 0.2702 \\
    \hline
\end{tabular}
\label{table_2}
\end{table*}

Table II lists the anti-detectability performances (values of $P_E$) of those approaches with different hiding payloads. From this table, there are several observations.

From this table, it is clear that the anti-detectability of the proposed generative steganographic approaches, i.e., S2IRT and SE-S2IRT, keeps at a high-level ($P_E$ is about 0.27) and generally outperform the other approaches significantly, when the hiding payload ranges from 0.5-4.0 \textit{bpp}. That is because the proposed generative steganographic approaches directly generate new images as stego-images without modification, and thus it is relatively hard for the steganalyzers to detect the presence of hidden information. In contrast, with the increase of hiding payload, the performances of the conventional steganographic approaches including S-UNIWARD, UT-6HPF-GAN, and Deep-Stego degrade significantly. In these approaches, as the secret message is hidden by modifying the existing cover images, hiding more information into the cover images leads to more distortion. Thus, it is much easier for the steganalyzers to successfully defeat these steganographic approaches, especially under high hiding payloads.

However, when the hiding payload is 0.1 \textit{bpp}, the $P_E$ values of the generative steganographic approaches, i.e., S2IRT, SE-S2IRT and SWE, are also about 0.27, which is lower than those of the conventional steganographic approaches, i.e., S-UNIWARD, UT-6HPF-GAN, and Deep-Stego. That is because the limited size of training dataset leads to the limited ability of generating realistic images of generative models, which makes some natural images to be distinguished from generated images correctly. However, this problem can be effectively alleviated by increasing the size of training dataset.

Also, it is notable that some $P_E$ values of the other steganographic approaches including S-UNIWARD, UT-6HPF-GAN, Deep-Stego, and SWE are empty or equal to 0 in Table I for the following reasons. The $P_E$ values of UT-6HPF-GAN and S-UNIWARD are empty when the payload is larger than 1.0 \textit{bpp}, because they cannot achieve the hiding capacity larger than 1.0 \textit{bpp}. Since Deep-Stego aims to hide a secret image within an image, they cannot hide the messages of small sizes and thus the $P_E$ values are empty when the hiding payload is smaller than 1.0. Moreover, it does not consider the resistance to steganalyzers, the stego-images can be easily detected by the steganalyzers, and thus the other $P_E$ values of Deep-Stego is equal to 0. Due to the irreversibility of DCGAN, SWE cannot extract the hidden information when the hiding payload is larger than 0.1 and thus the corresponding $P_E$ values of SWE are empty. Therefore, the high hiding capacities of Deep-Stego, SWE, S-UNIWARD and UT-6HPF-GAN come at the expense of security performance or information extraction accuracy.

\subsubsection{Imperceptibility}
Without the need for an existing cover image, the generative steganographic approaches, i.e., SWE, S2IRT and SE-S2IRT, directly transform a given secret message to a generated image for steganography. As there is no need of cover images in these approaches, the traditional reference image quality assessment strategies such as RSNR and SSIM are not suitable for evaluating the imperceptibility of these approaches. Thus, the typical model of reference less image quality assessment strategy, i.e., BRISQUE, is adopted in the experiments. However, BRISQUE is not sensitive to slight noise, and it is also experimentally found that the BRISQUE scores of the stego-images obtained by the conventional steganographic approaches are almost the same as those of original cover images, since those stego-images are obtained by slightly modifying the original cover images. Thus, BRISQUE is not suitable to evaluate the qualities of stego-images for the conventional steganographic approaches.

\begin{table}[]
\centering
\caption{The BRISQUE Scores of Stego-images of Those Approaches with Different Hiding Payloads}
\setlength{\tabcolsep}{3mm}
\begin{tabular}{ccccccc}
    \hline
    \multirow{2}{*}{Approaches} & \multicolumn{6}{c}{Hiding payloads (\textit{bpp})} \\ 
    \cline{2-7}
    & 0 & 0.1 & 0.5 & 1.0 & 2.0 & 4.0 \\
    \hline
    SWE & 9.07 & 8.69 & - & - & - & - \\
    S2IRT & 8.85 & 8.13 & 7.32 & 8.54 & 9.13 & 7.44 \\
    SE-S2IRT & 6.89 & 7.93 & 9.03 & 7.21 & 7.69 & 8.17 \\
    \hline
\end{tabular}
\label{table_1}
\end{table}

Therefore, we only compute and compare the BRISQUE scores for the stego-images generated by the generative steganographic approaches, i.e., SWE, S2IRT, and SE-S2IRT, with different hiding payloads. Table III shows the average BRISQUE scores of the stego-images generated by different approaches with different hiding payloads. Smaller BRISQUE score means higher image quality. From this table, although the BRISQUE scores of SWE are close to those of S2IRT and SE-S2IRT, SWE cannot accurately extract the hidden information when the hiding payload is higher than 0.1. Also, in S2IRT and SE-S2IRT, it is clear that the imperceptibility of the generated images without hiding message ($bpp=0$) is similar to the generated images with the hiding payload ranging from 0.1-4.0. That indicates the quality of stego-images generated by S2IRT and SE-S2IRT will not be affected with the increase of hiding capacity, mainly due to the efficiency and robustness of message encoding and the powerful ability of generating high-quality images of Glow model.

\begin{figure}[!t]
\centering
\subfloat[]{\includegraphics[width=1.7in]{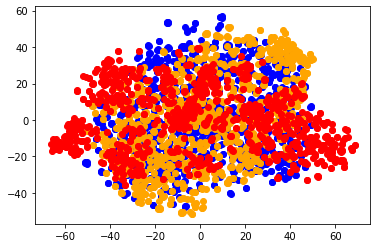}%
\label{fig_first_case}}
% \hfil
\subfloat[]{\includegraphics[width=1.7in]{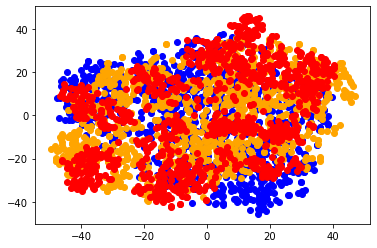}%
\label{fig_second_case}}
\caption{The t-SNE data of real images and the images generated by (a) S2IRT and (b) SE-S2IRT with and without information hiding. The data of real images, images generated without information hiding, and images generated with hiding payload of 4 \textit{bpp} is marked by blue, red, and yellow dots, respectively.}
\label{fig_6}
\end{figure}

To further illustrate imperceptibility of the proposed steganographic approaches, we also compute and compare the t-SNE data of the real images and the images generated with and without information hiding. Where, t-SNE is the dimensionality reduction strategy, which transforms high-dimensional image data to 2-dimensional data so as to observe the distribution of values of image data. From Fig. 6, it is clear that it is hard to distinguish the real images, the images generated without information hiding, and those with information hiding.

\begin{figure*}[!t]
\centering
\subfloat[]{\includegraphics[width=1.7in]{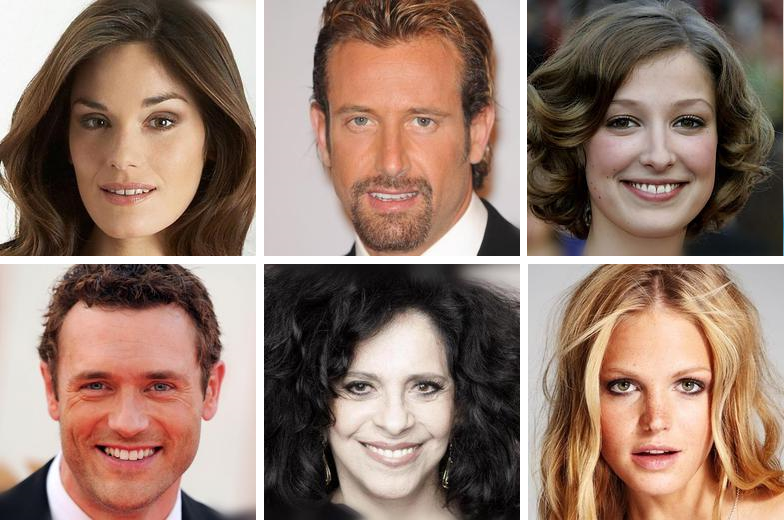}%
\label{fig_first_case}}
% \hfil
\subfloat[]{\includegraphics[width=1.7in]{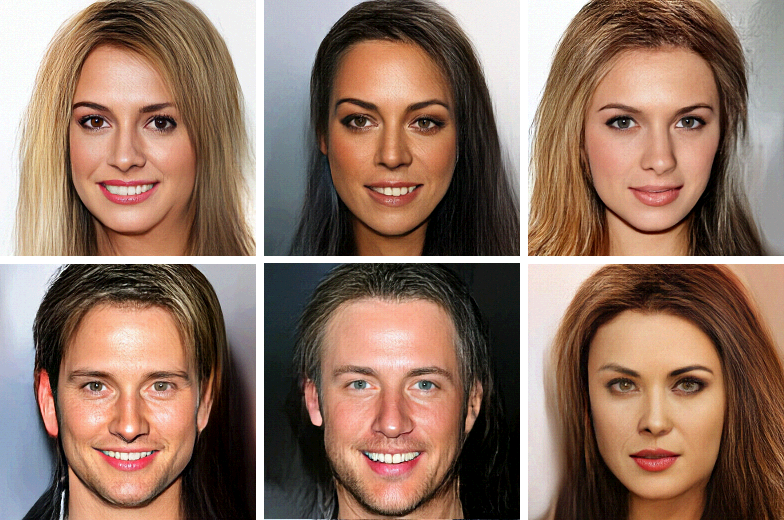}%
\label{fig7_second_case}}
\subfloat[]{\includegraphics[width=1.7in]{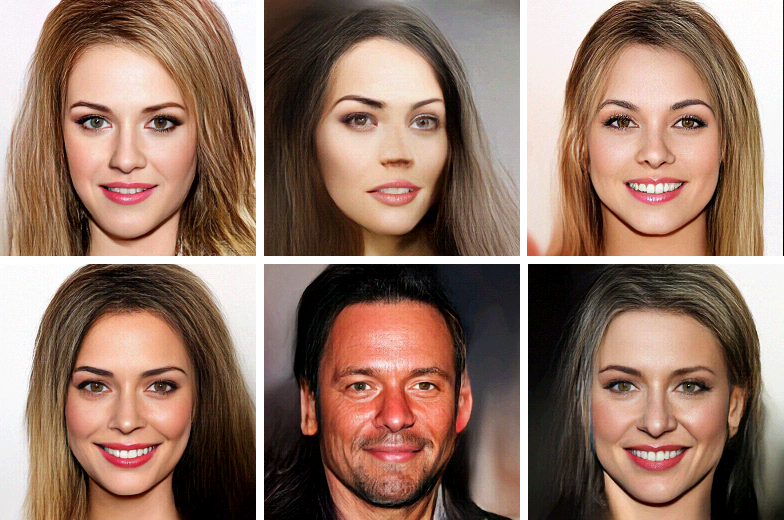}%
\label{fig7_third_case}}
\subfloat[]{\includegraphics[width=1.7in]{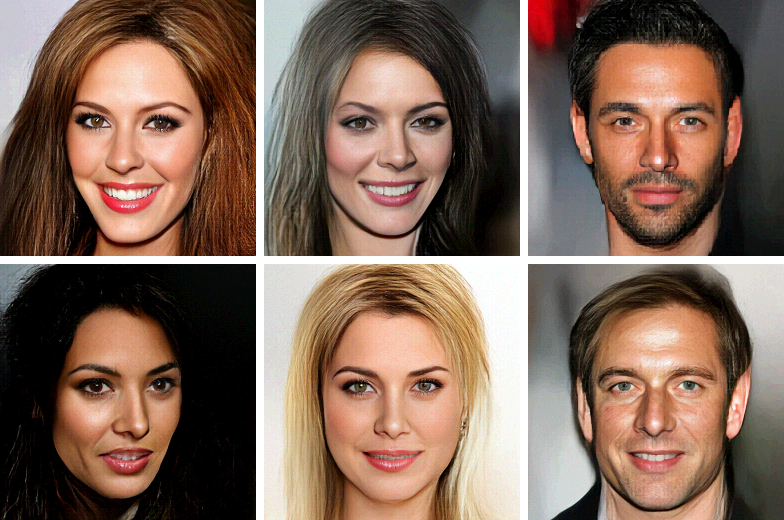}%
\label{fig7_fourth_case}}
\caption{The real images and generated face images. (a) The real face images, (b) the images generated by S2IRT without information hiding, (c) the images generated by S2IRT with hiding payload of 2 \textit{bpp}, and (d) the images generated by S2IRT with hiding payload of 4 \textit{bpp}.}
\label{fig_7}
\end{figure*}

Fig. 7 shows toy examples of real images and images generated by S2IRT without and with information hiding. It is clear that S2IRT can achieve excellent hiding capacity, i.e., up to 4.0 \textit{bpp}. Meanwhile, it is hard to distinguish the images generated with information hiding from the images generated without information hiding. Also, mainly due to the insufficient training of Glow model, it is notable that there are some slight visual differences between the real images and the generated images. The visual differences can be further reduced by training the Glow model on larger-sized datasets.

According to the above, the proposed steganographic approaches can consistently achieve desirable imperceptibility performances under different hiding payloads.

\begin{table}[]
\centering
\caption{The $IE_A$ Values of Stego-images of Those Proposed Approaches with Different Hiding Payloads}
\begin{tabular}{ccccc}
    \hline
    \multirow{2}{*}{Attack} & \multirow{2}{*}{Approaches} & \multicolumn{3}{c}{Hiding payloads (\textit{bpp})} \\ 
    \cline{3-5}
    & & 0.1 & 0.5 & 1.0\\
    \hline
    \multirow{2}{*}{Intensity change with 2\%} & S2IRT & 0.68 & 0.63 & 0.60 \\
    & SE-S2IRT & 1.00 & 0.91 & 0.75 \\
    \hline
    \multirow{2}{*}{Contrast enhancement with 5\%} & S2IRT & 0.66 & 0.62 & 0.59 \\
    & SE-S2IRT & 1.00 & 0.96 & 0.77 \\
    \hline
    \multirow{2}{*}{\makecell[c]{Salt \& Pepper Noise with \\ the factor of 3\%}} & S2IRT & 0.69 & 0.63 & 0.61 \\
    & SE-S2IRT & 0.91 & 0.82 & 0.73 \\
    \hline
    \multirow{2}{*}{\makecell[c]{Gaussian noise with standard \\ deviation $\sigma =10^{-3}$}} & S2IRT & 0.66 & 0.61 & 0.58 \\
    & SE-S2IRT & 0.84 & 0.78 & 0.69 \\
    \hline
    \multirow{2}{*}{JPEG compression 90} & S2IRT & 0.66 & 0.62 & 0.59 \\
    & SE-S2IRT & 0.80 & 0.73 & 0.61 \\
    \hline
    \multirow{2}{*}{Rotation $\pm 0.75^{\circ}$} & S2IRT & 0.66 & 0.60 & 0.58 \\
    & SE-S2IRT & 0.85 & 0.79 & 0.71 \\
    \hline
    \multirow{2}{*}{Image sterilization} & S2IRT & 1.00 & 0.95 & 0.61 \\
    & SE-S2IRT & 1.00 & 1.00 & 0.99 \\
    \hline
\end{tabular}
\label{table_4}
\end{table}

\subsubsection{Robustness}
To test and compare the robustness performances of S2IRT and SE-S2IRT, we adopt a set of common attacks including intensity change, contrast enhancement, salt \& pepper noise, Gaussian noise, JPEG compression, rotation as well as image sterilization \cite{r50}, which aims remove secret data hidden within an image. Then, these attacks are conducted on the generated stego-images. The accuracy rates of information extraction of the two approaches are listed in Table IV.

From Table IV, it can be clearly observed that SE-S2IRT performs better than S2IRT in the aspect of robustness for the following reasons. In SE-S2IRT, once the group \textit{No}. of one element arranged into one of K adjacent positions is changed, only the position choice \textit{Nos}. for arranging the elements into the K adjacent positions will be changed, and thus only $\sum^{K-1}_{i=1} \lfloor \log_2 (K-i+1) \rfloor$ bits will be affected at most. On the contrary, in S2IRT, once the group \textit{No}. of one element is changed, the position choice \textit{Nos}. for arranging the elements of most groups would be changed. Thus, most of the secret bits encoded as the position arrangement for these groups will be affected. Therefore, SE-S2IRT can improve the robustness performance of S2IRT significantly. Also, it is clear that SE-S2IRT achieves good robustness to intensity change, contrast enhancement and image sterilization, but relatively low robustness to Salt \& Pepper Noise, Gaussian noise, and rotation.

\begin{table}[]
\centering
\caption{The Probabilities of Successful Cracking Message Hidden by S2IRT and SE-S2IRT}
\setlength{\tabcolsep}{1mm}
\begin{tabular}{ccccc}
    \hline
    \multirow{2}{*}{K} & \multirow{2}{*}{Approaches} & \multicolumn{3}{c}{n} \\
    \cline{3-5}
    & & 10 & 20 & 30\\
    \hline
    \multirow{2}{*}{10} & S2IRT & $0.742\times 10^{-67}$ & $0.987\times 10^{-177}$ & $0.805\times 10^{-232}$ \\
    & SE-S2IRT & $0.637\times 10^{-57}$ & $0.406\times 10^{-114}$ & $0.102\times 10^{-142}$ \\
    \hline
    \multirow{2}{*}{20} & S2IRT & $0.322\times 10^{-231}$ & $0.209\times 10^{-490}$ & $0.396\times 10^{-614}$ \\
    & SE-S2IRT & $0.277\times 10^{-162}$ & $0.771\times 10^{-325}$ & $0.406\times 10^{-406}$ \\
    \hline
    \multirow{2}{*}{30} & S2IRT & $0.104\times 10^{-403}$ & $0.561\times 10^{-833}$ & $0.853\times 10^{-1043}$ \\
    & SE-S2IRT & $0.107\times 10^{-282}$ & $0.115\times 10^{-565}$ & $0.379\times 10^{-707}$ \\
    \hline
\end{tabular}
\label{table_5}
\end{table}

\subsection{Security Analysis}
We analyze the security of the proposed steganographic approaches in this subsection. Suppose an attacker can intercept the shared information of $K$ and $n$, but cannot obtain the $Key$, which is protected well by communication participants. According to Section 4.4 and 5.3, the probabilities of successfully cracking the secret messages hidden by S2IRT and SE-S2IRT can be computed by
\begin{equation}
\label{deqn_ex1a}
P_c = \frac{1}{\prod^{K-1}_{i=1}s_i}
\end{equation}
Where, $s_i=C(N-(i-1)n,n)$ in the S2IRT scheme and $s_i=[C(K-i+1,1)]^n$ in the SE-S2IRT scheme. The computed $P_c$ values with different $K$ and $n$ are listed in Table V. It is clear that the cracking probabilities of proposed steganographic approaches are very small, and thus they are secure enough against a malicious attack of random guess.

\section{Conclusion}
This paper has provided a novel idea for generative steganography. It presents two secret-to-image reversible transformation schemes, i.e., S2IRT and SE-S2IRT, based on the Glow model to implement generative steganography. The proposed steganographic approaches can achieve excellent hiding capacity (over 4 \textit{bpp}) and accurate information extraction (almost 100\% accuracy rate) while maintaining desirable anti-detectability and imperceptibility, and outperform the state-of-the-arts significantly. Moreover, SE-S2IRT can improve the robustness to a variety of common attacks significantly.

In many practical steganographic tasks, it is required to hide a large amount of information in an image while maintaining high extraction accuracy and desirable anti-detectability, imperceptibility and robustness. The pro-posed steganographic approaches can handle these tasks well. Hence, we can conclude that the proposed approaches have important practical significance in the area of information hiding. In future, we intend to extent the proposed approaches to generate other common types of images such as object images and landscape images for steganography.

\bibliographystyle{IEEEtran}
\bibliography{ref}

\newpage
 

% \end{IEEEbiography}



\begin{IEEEbiography}[{\includegraphics[width=1in,height=1.25in,clip,keepaspectratio]{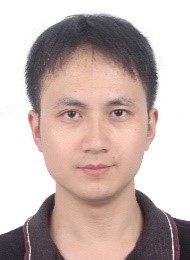}}]{Zhili Zhou}
(M’19) received his MS and PhD degrees in Computer Application at the School of Information Science and Engineering from Hunan University, in 2010 and 2014, respectively. He is currently a professor with School of Computer and Software, Nanjing University of Information Science and Technology, China. Also, he was a Postdoctoral Fellow with the Department of Electrical and Computer Engineering, University of Windsor, Canada. His current research interests include Multimedia Security, In-formation Hiding, Digital Forensics, Blockchain, and Secret Sharing. He is serving as an Associate Editor of Journal of Real-Time Image Processing and Security and Communication Networks.
\end{IEEEbiography}

\begin{IEEEbiography}[{\includegraphics[width=1in,height=1.25in,clip,keepaspectratio]{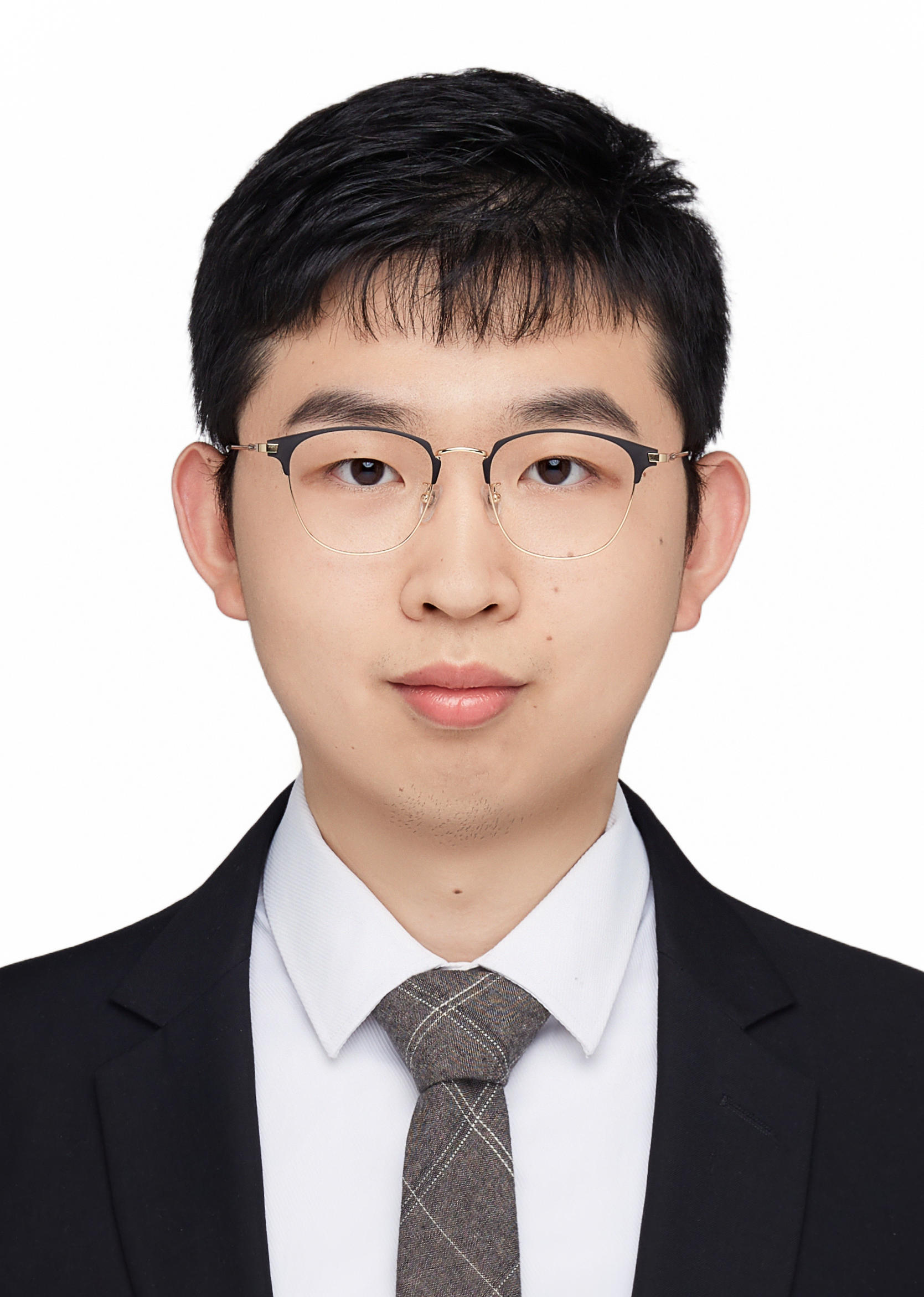}}]{Yuecheng Su}
is a graduate student. He is currently pursuing his MS degree in Department of Computer and Software, Nanjing University of Information Science and Technology, China. His current re-search interests include information hiding and digital forensics.
\end{IEEEbiography}

\begin{IEEEbiography}[{\includegraphics[width=1in,height=1.25in,clip,keepaspectratio]{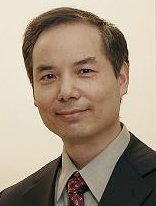}}]{Q. M. Jonathan Wu}
(M’92-SM’09) received the Ph.D. degree in electrical engineering from the University of Wales, Swansea, United Kingdom, in 1990. He was affiliated with the National Research Council of Canada for ten years beginning, in 1995, where he became a senior research officer and a group leader. He is currently a professor with the Department of Electrical and Computer Engineering, University of Windsor, Windsor, Ontario, Canada. He has published more than 300 peer-reviewed papers in computer vision, image processing, intelligent systems, robotics, and integrated microsystems. His current research interests include 3-D computer vision, active video object tracking and extraction, interactive multimedia, sensor analysis and fusion, and visual sensor networks. Dr. Wu held the Tier 1 Canada Research Chair in Automotive Sensors and Information Systems since 2005. He is an associate editor for the IEEE Trans-action on Cybernetics, the IEEE Transactions on Circuits and Systems for Video Technology,  the Journal of Cognitive Computation, and the Neurocomputing. He has served on technical program committees and international advisory committees for many prestigious conferences. He is a Fellow of Canadian Academy of Engineering.
\end{IEEEbiography}

\begin{IEEEbiography}[{\includegraphics[width=1in,height=1.25in,clip,keepaspectratio]{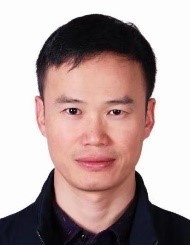}}]{Zhangjie Fu}
received the Ph.D. degree in computer science from the College of Computer, Hunan University, China, in 2012. He is currently a professor in the School of Computer, Nanjing University of Information Science and Technology, China. His research interests include cloud and outsourcing security, digital forensics, network and information security. His research has been supported by NSFC, PAPD, and GYHY. He is a member of the IEEE, and a member of ACM.
\end{IEEEbiography}

\begin{IEEEbiography}[{\includegraphics[width=1in,height=1.25in,clip,keepaspectratio]{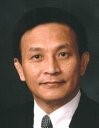}}]{Yun-Qing Shi}
(Life Fellow, IEEE) received the M.S. degree from Shanghai Jiao Tong University, China, and the Ph.D. degree from the University of Pittsburgh, USA. In 1987, he joined the New Jersey Institute of Technology, USA. He is the author or coauthor of more than 400 articles, one book, five book chapters, and an editor of ten books. He holds 30 U.S. patents. His research interests include data hiding, forensics, information assurance, and visual signal processing and communications. He has been a member of a few IEEE technical committees and a fellow of the National Academy of Inventors (NAI) since 2017. He has served as the Technical Program Chair for the IEEE ICME07, the Co-Technical Chair for the IEEE MMSP05, as well as IWDW since 2006, the Co-General Chair for the IEEE MMSP02, and a Distinguished Lecturer for the IEEE CASS. He has served as an Associate Editor for the IEEE Transactions on Signal Processing, the IEEE Transactions on Information Forensics and Security, and the IEEE Transactions on Circuits and Systems and an editorial board member for few journals.
\end{IEEEbiography}

\vfill

\end{document}